\documentclass[fleqn,12pt,twoside]{article}
\usepackage{espcrc1}

\usepackage{graphicx}

\newcommand{\AmS}{{\protect\the\textfont2
  A\kern-.1667em\lower.5ex\hbox{M}\kern-.125emS}}

\title{Direct photons and thermal dileptons: A theoretical review}

\author{Charles Gale \address{Department of Physics, McGill University,\\
3600 University St., Montreal, QC, H3A 2T8, Canada}}
       
\begin{document}

\maketitle

\begin{abstract}
We discuss the measurement of electromagnetic radiation produced in
heavy ion collisions at SPS energies. We review the low invariant 
mass dilepton
sector, the real photon data, and the spectra of intermediate mass 
dimuons. Along with this, we discuss the theoretical interpretations 
of those observables.  
\end{abstract}

\section{Introduction}

The measurement of electromagnetic probes provides a valuable and a
necessary complement to that of hadronic observables in an environment
as potentially complex as that of ultrarelativistic heavy ion
collisions. This fact owes partly to the weakness of final-state
interactions in photon emission (real or virtual) but is also related
to the richness of the information carried by the electromagnetic
observables. Those statements can be made quantitative in the
following. In a finite-temperature strongly-interacting medium the 
differential rate for real photon emission is given by
\cite{rate,gk} 

\begin{equation}
\omega {{d^3 R}\over{d^3 k}}\ =\ - {{1}\over{(2 \pi)^3}}\, {\rm Im}\,\Pi^{\mu}_{\mu}
\, {{1}\over{e^{\beta \omega} - 1}}\ ,
\label{eq1}
\end{equation}
while that for virtual photon emission is 
\begin{equation}
E_+ E_- {{d^6 R}\over{d^3 p_+ d^3 p_-}}\ = \ {{2 e^2}\over{(2
\pi)^6}}\,{{1}\over{k^4}} L^{\mu \nu} {\rm Im}\,\Pi_{\mu \nu}\, 
{{1}\over{e^{\beta \omega} - 1}}\ .
\end{equation}
In the above equations $\omega$ is the photon energy, $\beta$ is
the inverse temperature, $k^2$ is the virtual photon invariant mass,
$L_{\mu \nu}$ is a lepton tensor, 
and the $E$'s and $p$'s are the lepton energy and momenta. 
$\Pi_{\mu \nu}$ is the retarded, finite-temperature photon self-energy:
a direct window to the many-body physics. 
Vector Meson Dominance (VMD) couples the photon field (real or virtual)
to hadronic matter in the confined sector via the current-field 
identity \cite{vmd}: $J^\mu = \sum_i g_i \, \phi^\mu_i$, where the sum
runs over the vector meson fields $\phi^\mu_i$.  In the
sector where the relevant degrees of freedom are partonic: $J^\mu =
\sum_q e_q \bar{\psi}_q \gamma^\mu \psi_q$, where $e_q$ is a quark
charge. In the region where
confinement is effective, the dilepton production rate can then be
directly related to the in-medium spectral density of the vector mesons.
There lies one of the reasons for measuring and understanding the dilepton
spectrum: the vector and axial vector correlators will mix
via interactions with the thermal background \cite{dei}. In the limit
of chiral symmetry restoration, those correlators are expected to
become degenerate \cite{ks}. To obtain information on the axial
current-current correlator through the measurement of electromagnetic
radiation constitutes a challenge to which we will return later. 
Also note that the importance of real photons and dileptons as tools for
quark-gluon plasma (QGP) diagnostics was realized early on
\cite{fein,es} and has
fueled the interest in those variables ever since.

As RHIC begins its operation, it is sensible to assess the situation in
terms of what has been measured and learnt so far at SPS energies. 
We shall discuss in
turn the low dilepton invariant mass region, real photon measurements,
and the intermediate dilepton invariant mass sector. It is futile   
here to attempt to cover and do justice to the many works that are 
relevant to
this discussion. Also, in the spirit of a
review with length constraints, we point to the appropriate references for
the technically-intensive issues.

\section{Low mass dileptons}

\begin{figure}[t!]
\begin{center}
\includegraphics[height=9.0cm]{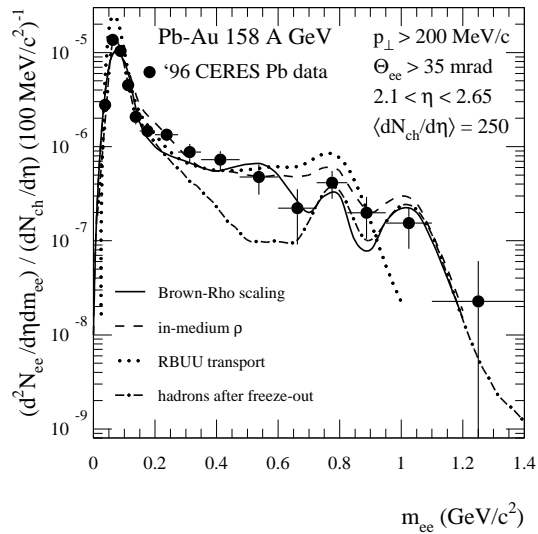}
\end{center}
\vspace*{-1cm}
\caption{The low invariant mass CERES data with three theoretical
curves representing models that are described in the text. The 
figure is from \protect\cite{itzhak}.  \label{lowmass}}
\end{figure}

This kinematical region, first explored in HELIOS dimuon measurements
\cite{helios}, was scrutinized by CERES in a series of dielectron
measurements which have generated a considerable amount of theoretical
activity \cite{ceres,ceresqm}. The experimental measurements, together
with some of their theoretical interpretations are shown in 
Fig.~\ref{lowmass}. As the status of the theoretical understanding of
those inspiring data has not been profoundly modified since our
last meeting in this series, there is no need here to go into a great
deal of detail. The Brown-Rho scaling hypothesis is described in a
recent review \cite{br}. The curve labeled ``in-medium $\rho$''
reflects an in-medium calculation of the vector meson spectral density
\cite{rw}, which is directly related to the dilepton emission rate
through the current-field identity \cite{gk}. The results of a
transport calculation involving an incoherent sum of meson reaction is
also shown \cite{vk}. The ``in-medium'' calculation is a representation
of fairly conventional many-body physics, as the vector mesons are
broadened by interactions with nucleons and created mesons, as well as
with the pions in the pion cloud. It turns out the the pole mass of the
$\rho$ is not drastically modified by those interactions, owing to 
cancellations between  
channels \cite{rw,rg}. QCD sum rules can only provide a broad
consistency-check of some of the phenomenological models \cite{srule}.   
An experimental attempt to distinguish between
the approaches listed above involves high  resolution measurements
at and around the vector meson peaks. The calculations relying on
broadened spectral densities are lower there because of unitarity
requirements \cite{rg,cb}. Also, a study of the low mass dilepton excess as
a function of the beam energy and of the baryon density would also
constrain the models: runs at higher baryon densities may highlight
novel many-body manifestations such as mixing effects \cite{ot}. 
Those important refinements of the data would help to bring this exciting
chapter to some closure.

\section{Real photons}

The interest in the thermal emission of real photons partly stems from
the seminal suggestion by Feinberg that thermal 
electromagnetic emission might be an important process when a large
multiplicity of particles are produced in the final state \cite{fein}. 
To lowest order in the coupling constants, the fact that the photon
spectrum carries with it information about the electromagnetic
current-current correlator can be rewritten in the language of
relativistic kinetic theory. In a quark-gluon plasma, the production of
real photons will proceed through annihilation ($q \bar{q} \to \gamma g$)
and Compton ($g q \to \gamma q$, $g \bar{q} \to \gamma \bar{q}$) channels.
Therefore, the photons produced in the plasma carry information on the
thermodynamic state of the parton medium at the moment of their
production. However from a practical point of view, the 
signal-to-background ratio is smaller than that for dileptons by
roughly two orders  of magnitude, due especially to the large $\pi^0$
and $\eta$ decay contributions. This fact makes the real photon
measurements particularly challenging. At SPS energies, the first
attempts to observe direct photon production in ultrarelativistic
nuclear collisions with O and S beams found no significant excess
\cite{noexcess}.  The WA80 collaboration has obtained upper bounds
for the $p_T$ spectrum of real photons generated in S + Au
collisions at 200 AGeV \cite{wa80}. A measurement of 
direct photons in
$^{208}$Pb + $^{208}$Pb collisions at 158 AGeV was finally reported by 
WA98
\cite{wa98}.  
\begin{figure}[t!]
\begin{center}
\includegraphics[width=8.0cm,angle=0]{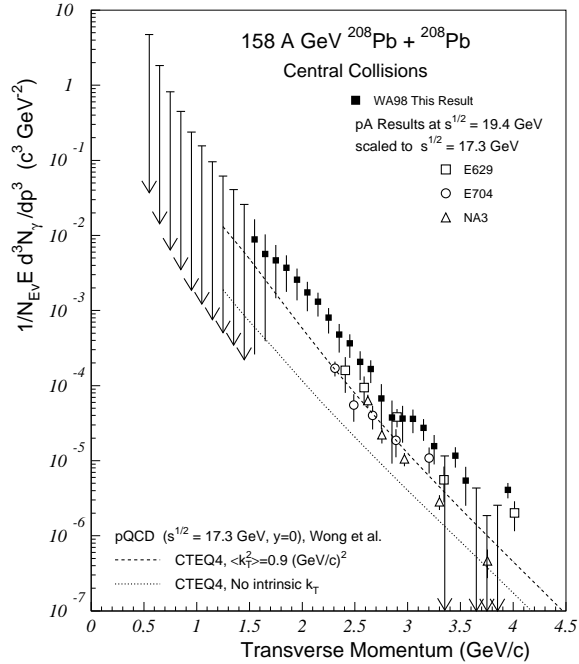}
\end{center}
\vspace*{-1cm}
\caption{The WA98 real photon invariant cross section data, as a 
function of photon transverse momentum \cite{wa98}. 
The pQCD calculation with intrinsic $k_\perp$ is from \cite{wong}.}  
\label{wa98}
\end{figure}

Recent times have witnessed a flurry of theoretical activity centered
on the theoretical interpretation of the WA98 photon measurements. The
data along with a pQCD estimate are shown in
Fig.~\ref{wa98}.  It is impossible to discuss here the many
calculations, but 
ingredients of some theoretical interpretations are not free 
from ambiguities and
those should be pointed out in order to make progress. Let us first
consider the high transverse momentum part of the measured
photon spectrum.  In hadron-hadron collisions at ISR energies, 
comparisons of   the measured high
$p_T$ sector with NLO-QCD direct photon 
calculations reveal that pQCD contributions are quantitatively
important \cite{pqcd1}. This component of the
spectrum should reflect the parton distribution functions. However, 
despite many years of experimental and theoretical
effort, the process of inclusive photon production in the collision of
elementary hadrons is not yet fully understood, and those uncertainties
propagate to nuclear collisions. A recent critical survey
concludes that the present fixed target data on inclusive prompt photon 
data at ISR energies ($\sqrt{s} \geq 23$ GeV) are inconsistent 
\cite{aur99}, making it
difficult to extract from them physically meaningful values of model
parameters. Bearing those
uncertainties in mind, the idea of attributing the pattern of
deviations between the measured direct photon cross sections and NLO
calculations to parton transverse momentum (neglected in
the NLO calculations) has however gathered support \cite{hus95,apan99}. Taking
a confinement radius of 0.5 fm, the intrinsic transverse
momentum generated thusly is $<k_\perp^2 > \simeq 0.32$ (GeV/c)$^2$.  
Furthermore,  soft-gluon emission will make this value larger 
\cite{css}.  The
resummation calculation for multiple soft-gluon emission in direct
photon production is challenging, and part of their contribution will
receive a non-perturbative component at the $\sqrt{s}$ values that are
relevant for this discussion. As a temporary substitute for a rigorous
calculation, effects of soft-gluon radiation are approximated by a
convolution of the LO cross section with a $k_\perp$-smearing function.
Early estimates obtained  $<k_\perp^2 > \simeq 0.9$ (GeV/c)$^2$ at
$\sqrt{s}$ = 20 GeV \cite{owens}. 

In addition to these considerations
(and to their inherent uncertainties), multiple initial state scattering
(the Cronin effect) in pA and AA collisions  will add to the parton 
transverse momenta \cite{wang,plf}, prior to the photon-forming interaction.  
Clearly, before a consensus is achieved, systematic studies of hadron
and photon spectra are needed in pp, pA, and AA collisions {\it at the
same energies.} This has not been possible in the past, and the necessary
energy-scaling procedures add to the ambiguities. This situation will not be
repeated at RHIC, where all three types of collisions will be
accessible to a given experimental setup. One may then ask how crucial
are the high temperatures advocated by early hydrodynamical
calculations \cite{dks}? As long as the quantitative role played by
parton momentum broadening in nuclear collisions is not fully
elucidated, it appears difficult to bring this issue to closure. The
global survey suggested above would go a long way in this direction. A
closely related issue is the use in the deconfined-QCD sector of the
photon production rates obtained from two-loop topologies at
finite-temperature \cite{dks,qcd1}. Those utilize the hard thermal loop
resummation method in finite temperature perturbation theory
\cite{htl}. This technique allows the investigation of screening
effects in a systematic way. At the one-loop level, this approach was
successfully used to compute the emission rate of hard real 
photons  \cite{bnnr,kls} and dileptons \cite{htldilep}.
The two-loop rates receive a large
boost from a colinear singularity, making the process labeled
``annihilation-with-scattering'' dominant \cite{qcd1}. Some progress has
been made in understanding those results in terms of an ordering of the
different length scales in the problem \cite{francois}. However, even
including obvious coupling constants scaling, it is
fair to say that the use at the SPS of the photon rates calculated
for asympotically high energies rests  on the belief that the
relative contributions to the photon spectrum of the different 
self-energy cuts are generic.
This still remains to be put on a firm theoretical foundation. 

The low $p_T$ part of the direct photon spectrum is linked to softer
physical contributions. Those typically will involve interactions
between degrees of freedom from the lower temperature 
confined sector. 
Let's consider the kinetic-theory representation of the 
hadronic reactions responsible for direct photon production. Those 
rates have been evaluated
first in Ref. \cite{kls}, have been verified subsequently 
\cite{sinha98,martine}, and involve a gas of $\pi$, $\eta$, $\rho$, and
$\omega$ mesons. Also, contributions involving the $a_1$ have been
shown to be quantitatively important \cite{a1}. Herein lies an ambiguity
in the hadronic phase: there is no {\it unique} way to implement a
chirally-symmetric model with vector mesons. A consequence of this is that a 
survey of the literature
rapidly reveals that photon production rates in a meson gas can differ 
by factors of
$\sim$ 3, at temperatures and transverse momenta that are germane to the
conditions discussed here \cite{a1,photons}. We advocate again a systematic
treatment of all the available hadronic data in order to reduce the
degeneracies in parameter-space of the effective theories involved
\cite{gg}. Furthermore, definite conclusions about the nature of the 
sources in the soft
part of the photon spectrum produced in heavy ion collisions are 
hindered by the following fact. Consider for a moment the WA80 photon
data (S + Au, at 200 AGeV), together with photon rates calculated with
the same vector meson spectral density as that used to interpret the
CERES data. The photon spectrum one obtains after a
simple fireball space-time evolution constitutes at least $\sim$ 16 \% of
the signal at $p_T$ = 1.5 GeV/c \cite{rw}. It is
known that the in-medium effects in this energy regime are largely 
baryon-driven \cite{rw}.  
This begs the question of how much is the corresponding contribution
to the WA98 Pb + Pb data, as the appropriate calculation has not yet
been done. There, an examination of the data with a LO-pQCD
calculation with an intrinsic $<k_\perp^2> \approx$ 1 (GeV/c)$^2$
yields $\sim$ 45\% of the measured yield at $p_T$ = 2.5 GeV/c
\cite{wong}. Barring the Cronin effect, the 
sources needing to be added then are
the contribution from the in-medium vector meson spectral densities and
the additional  contributions from the meson gas that are not 
included in them as, in self-energy parlance, they are specifically
two-loop contributions. This course of action is then laid out and preliminary 
results are promising \cite{hgr}. 
\begin{figure}[t!]
\begin{center}
\includegraphics[width=8.0cm,angle=0]{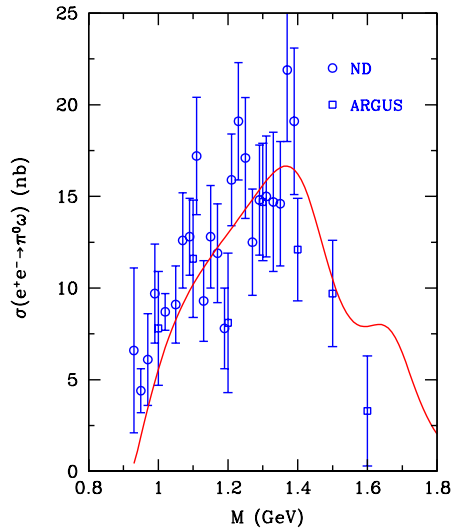}
\end{center}
\vspace*{-1cm}
\caption{The cross section for $e^+ e^- \to \pi^0 \omega$ as a function
of dilepton invariant mass. The solid
curve shown here is based on the model of 
\protect\cite{dol91}. The experimental
data are from the ND \cite{dol91} and ARGUS \cite{argus} collaborations.}
\label{eedata}
\end{figure}

Is therefore appears that uncertainties remain in the theoretical
interpretation of the direct photon spectrum, both in the low and high
transverse momentum regions. However, several of those difficulties can
be resolved. 

\begin{figure}[t!]
\begin{center}
\includegraphics[width=9.0cm,angle=-90]{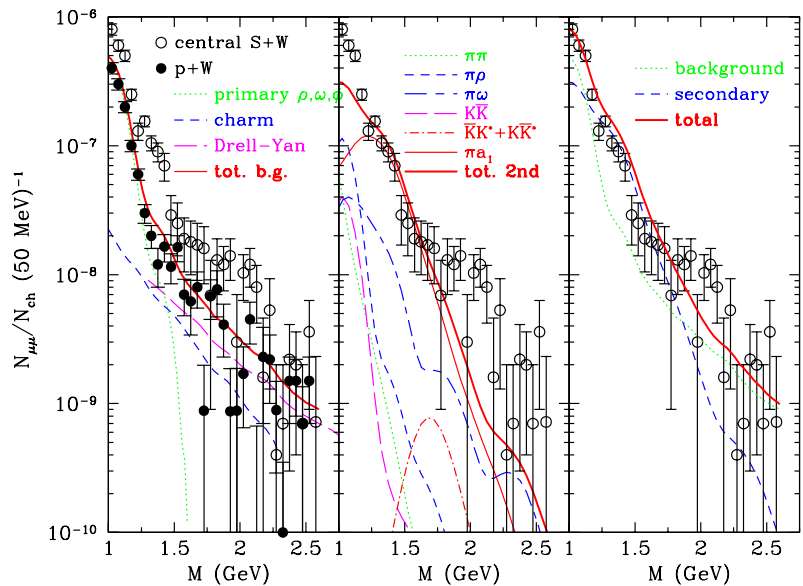}
\end{center}
\vspace*{-1cm}
\caption{Left panel: comparison of background, Drell-Yan,  and open
charm
decays with p+W and S+W Helios-3 dimuon data. Middle panel: some 
mesonic reactions 
contributing to lepton pair final states are shown. Right panel: the
sum, the background, and the secondary contributions are shown with the
data from central S+W collisions.} 
\label{helios3}
\end{figure}

\section{Intermediate mass dileptons}

Dimuon pairs in the intermediate invariant mass region ($m_\phi < M <
m_{J/\psi}$) have been measured at CERN by the Helios-3 \cite{h3} and 
NA38/NA50 collaborations \cite{naimr}. We consider those in turn. The
Helios-3 data is concerned with dimuon production in p + W and S + W
collisions at 200 AGeV. As in the low mass sector, an analysis 
of the heavy ion
data and of the proton-nucleus data revealed an excess of lepton pairs
over that expected from known sources; those being the direct decays of
primary vector mesons, the Drell-Yan contribution and the correlated
semileptonic decay of open charm mesons. Naturally, one should ask
whether there exist mesonic channels similar to the ones at play in
the low invariant mass region that could significantly contribute in
this kinematical region. 
However a precise determination of the specific
channels starting from  effective chiral Lagrangian techniques  remained
elusive, as in this invariant region the basic VMD form factors will be
significantly off-shell and  there is enough  phase space for
the initial channels to couple to a variety of high mass vector mesons
\cite{pdg}. 
Fortunately, there exists a wealth of data concerning $e^+
e^- \to$ hadrons in the appropriate invariant mass range
\cite{dol91}. Those data can be analyzed channel-by-channel
\cite{ioulia}: an example appears in Fig.~\ref{eedata}.
Inverting the channels that have two-body mesonic final states, a reaction 
database can be constructed. Those contributions can then be used in
a transport approach that can be directly compared with experimental
measurements. The procedure described here was used in conjunction with
a cascade-type model and compared with the Helios-3 data \cite{ligale},
and some results are shown in Fig.~\ref{helios3}. 
\begin{figure}[t!]
\vspace*{-1cm}
\begin{flushleft}
\includegraphics[width=6.5cm,angle=-90]{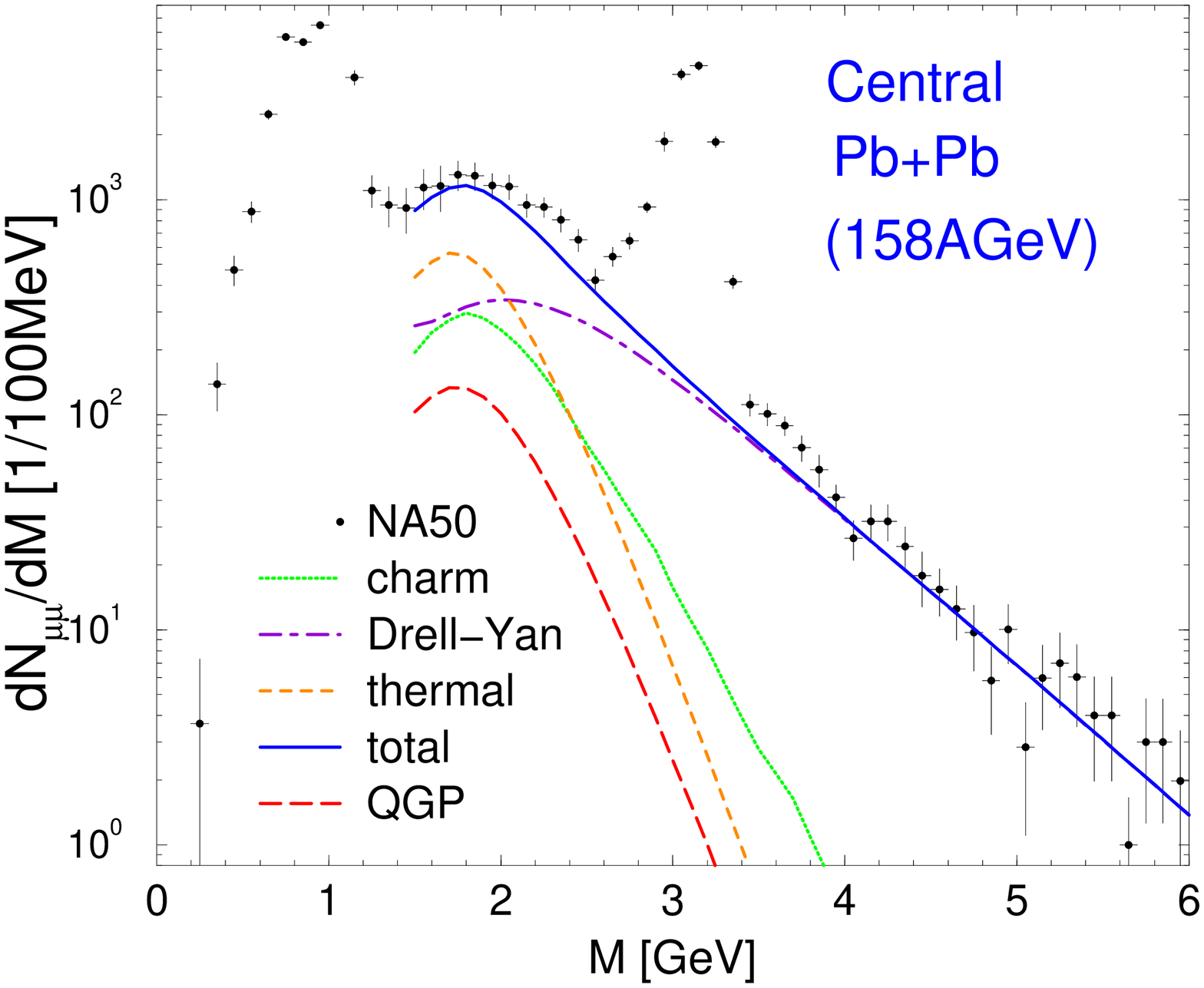}
\end{flushleft}
\end{figure}
\begin{figure}[t!]
\vspace*{-6.93cm}
\begin{flushright}
\includegraphics[width=5.5cm,angle=90]{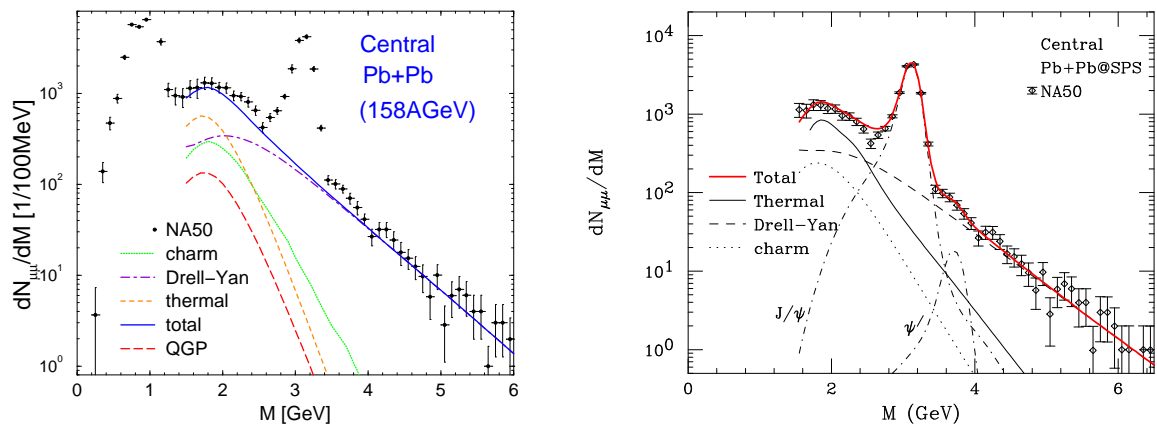}
\end{flushright}
\vspace*{-1cm}
\caption{Comparison with the NA50 intermediate mass dimuon data. The left
panel is a calculation from Ref.~\cite{rs}. The right panel is a
calculation from Ref.~\cite{kgss}. There, the ``thermal'' contribution
includes the QGP component.}
\label{na50}
\end{figure}
It appears that once the ``thermal'' channels are included, the
measured observable calls for little or no additional sources. Recently, this
approach was applied to the intermediate mass dimuon data measured by
NA50 \cite{rs}. A result of this calculation is shown in
Fig.~\ref{na50}, together with the result of a similar approach. 
Those two calculations differ mainly in the dynamical models used to
integrate out the dilepton rates, and in their quark-gluon plasma
content. Whereas a study is underway 
\cite{kgss2} to try and
pin down, as much as the data allows, model parameters such as the initial 
temperature (while
still being consistent with the global set of measured observables at
those bombarding energies), an important message emerges:
the importance of the secondary mesonic reactions \cite{ligale} can not 
be denied. To completely rule out a mechanism like $c \bar{c}$
enhancement, for example, a better measurement is needed and a
dedicated experiment is in the planning stages \cite{carlos}.

\section{Conclusion}

The measurement of electromagnetic radiation at the SPS has generated
exciting data that has considerably motivated and stirred the field as
a whole. It is appropriate to pause at this point and to reflect on
what has been learnt.  The low mass dileptons show strong indications of the
signature of a genuine, in-medium many-body effect. This is an
achievement that should not be underplayed.  As elaborated on
in the text, the theoretical effort on this is ongoing and high
resolution data at the vector meson masses will bring this discussion
to a new level of quantitative assessment.  In the real photon sector,
a systematic evaluation of the nuclear $k_\perp$ broadening in hadron
and electromagnetic spectra is called for. Similarly, the theoretical 
framework used for real photons has to be consistent, and in fact
identical,  with that used in
the virtual photon domain. This is vital for the sake of theoretical
consistency and also in view of the fact that those data do show 
common global features \cite{kampfer}.  
Some issues discussed here represent well-posed problems and
answers are to be expected soon. 

Have we seen a restoration (even
partial) of chiral
symmetry? Without a window on the axial correlator this issue remains
inconclusive. This will be difficult to assess from studies of low 
mass dileptons as the axial vector channel is far below those of
stronger sources \cite{vk,rg}. However, we suggest that it might 
be possible in real
photon and/or intermediate mass dilepton measurements, as there the
contribution of the $a_1$, for example, is not small \cite{a1,ligale}.
Before this can be done, the calculations of the axial vector spectral 
density have to be brought to a level of sophistication close to those
that exist in the vector channel. Have we seen a quark-gluon plasma? 
As it is, some analyses of the hard portion of the photon spectrum and
of the intermediate mass dilepton data do contain a QGP component,
albeit at different levels. In my opinion is it still too early to
tell, but keeping in mind the caveats outlined in
this talk, it is fair to say that this carries the suggestion that the
SPS has ventured in a threshold energy density region. RHIC has
the tantalizing mandate to go deeper in this new realm.

\section*{Acknowledgments}
It is a pleasure for me to thank my collaborators.
This work is supported in part by the  Natural Sciences and 
Engineering Research Council of Canada, and in part by the
Fonds FCAR of the Quebec Government.


\begin{thebibliography}{99}
\bibitem{rate} L. D. McLerran and T. Toimela, Phys. Rev. D {\bf 31}
(1985) 1545; H. A. Weldon, Phys. Rev. D {\bf 42} (1990) 2384.
\bibitem{gk} Charles Gale and Joseph I. Kapusta,  Nucl. Phys. 
{\bf B357} (1991) 65
\bibitem{vmd}J. J. Sakurai, {\it Currents and Mesons}, (University of
Chicago Press, Chicago, 1969); H. B. O'Connell, B. C. Pearse, A. W.
Thomas, and A. G. Williams, Prog. Part. Nucl. Phys. {\bf 39} (1997)
201.
\bibitem{dei}M. Dey, V. Eletsky and B. I. Ioffe, Phys. Lett. {\bf B252}
(1990) 620. 
\bibitem{ks}J. I. Kapusta and E. V. Shuryak, Phys. Rev. D {\bf 49}
(1994) 4694.
\bibitem{fein}E. L. Feinberg, Nuovo Cimento {\bf 34} A (1976) 391.
\bibitem{es}E. V. Shuryak, Phys. Lett. {\bf B78} (1978) 150.
\bibitem{helios}A. L. S. Angelis, Eur. Phys. J. C {\bf 13} (2000)
433, and references therein.
\bibitem{ceres}B. Lenkeit for the CERES collaboration, Nucl. Phys. {\bf
A661} (1999) 23c, and references therein.
\bibitem{ceresqm}H. Appelshaeuser, these proceedings.
\bibitem{br}G. E. Brown and Mannque Rho, hep-ph/0103102, and references
therein.
\bibitem{rw}R. Rapp and J. Wambach, Adv. Nuc. Phys. {\bf 25} (2000) 1 
[hep-ph/9909229].
\bibitem{itzhak} I. Tserruya, nucl-ex/9912003.
\bibitem{vk}V. Koch and C. Song, Phys. Rev. C {\bf 36} (1996) 1903.
\bibitem{rg}Ralf Rapp and Charles Gale, Phys. Rev. C {\bf 60} (1999) 
024903.
\bibitem{srule}S. Leupold, W. Peters, and U. Mosel, Nucl. Phys. {\bf
A628} (1998) 311.
\bibitem{cb}W. Cassing and E. L. Bratkovskaya, Phys. Rep. {\bf 308}
(1999) 65.
\bibitem{ot}G. Wolf, B. Friman, and M. Soyeur, Nucl. Phys. {\bf A640}
(1998) 129; O. Teodorescu, A. K. Dutt-Mazumder, and C. Gale,
Phys. Rev. C {\bf 61} (2000) 051901; Phys. Rev. C {\bf 63} (2001) 034903.
\bibitem{noexcess}T. {\AA}kesson {\it et al.}, Z. Phys. C {\bf 46}
(1990) 369; R. Albrecht {\it et al.}, Z. Phys C {\bf 51} (1991) 1; R.
Baur {\it et al.}, Z. Phys. C {\bf 71} (1996) 571.
\bibitem{wa80} R. Albrecht {\it et al.}, Phys. Rev. Lett. {\bf 76} 
(1996) 3506. 
\bibitem{wa98}M. M. Agarwal {\it et al.}, Phys. Rev. Lett. {\bf 85}
(2000) 3595; nucl-ex/0006007; A. Lebedev, these proceedings.
\bibitem{wong}C. Y. Wong and H. Wang, Phys. Rev. C {\bf 58} (1998) 376.
\bibitem{pqcd1}P. Aurenche, R. Baier, M. Fontannaz, and D.
Schiff, Nucl. Phys. {\bf B286} (1987) 509; Nucl. Phys. {\bf B297}
(1988) 661. 
\bibitem{aur99}P. Aurenche {\it et al.}, Eur. Phys. J. C {\bf 9} (1999)
107.
\bibitem{hus95}J. Huston {\it et al.}, Phys. Rev. D {\bf 51} (1995)
6139.
\bibitem{apan99}L. Apanasevich {\it et al.}, Phys. Rev. D {\bf 59}
(1999) 074007.
\bibitem{css}J. C. Collins, D. E. Soper, and G. Sterman, Nucl. Phys.
{\bf B250} (1985) 199.
\bibitem{owens}J. Owens, Rev. Mod. Phys. {\bf 59} (1987) 465.
\bibitem{wang}X. N. Wang, Phys. Rev. C {\bf 61} (2000) 064910.
\bibitem{plf}G. Papp, P. Levai, and G. Fai, Phys. Rev. C {\bf 61}
(2000) 021902; G. Papp {\it et al.}, these proceedings.
\bibitem{dks}Dinesh Kumar Srivastava and Bikash Sinha, nucl-th/0006018.
\bibitem{qcd1}P. Aurenche, F. G\'elis, R. Kobes, H. Zaraket, Phys.
Rev. D {\bf 58} (1998) 085003. 
\bibitem{htl}R. D. Pisarski, Nucl. Phys. {\bf B309} (1988) 476; Phys.
Rev. Lett. {\bf 63} (1989) 1129; Nucl. Phys. {\bf A525} (1991) 175c; E.
Braaten and R. Pisarski, Phys. Rev. Lett. {\bf 64} (1990) 1338; Nucl.
Phys. {\bf B337} (1990) 569; Nucl. Phys. {\bf B339} (1990) 310; J.
Frenkel and J. C. Taylor, Nucl. Phys. {\bf B334} (1990) 199.
\bibitem{bnnr}R. Baier, H. Nakkagawa, A. Ni\'egawa, and K. Redlich, Z.
Phys. C {\bf 53} (1992) 433.
\bibitem{kls}Joseph Kapusta, Peter Lichard, and David Seibert, Phys.
Rev. D {\bf 44} (1991) 2774.
\bibitem{htldilep}E. Braaten, R. Pisarski, and T. C. Yuan, Phys. Rev.
Lett. {\bf 64} (1990) 2242; S. M. H. Wong, Z. Phys. C {\bf 53} (1992)
465; T. Altherr and P. V. Ruuskanen, Nucl. Phys.
{\bf B380} (1992) 377; 
Markus H. Thoma and Christoph T. Traxler, 
Phys. Rev. D {\bf 56} (1997) 198. 
\bibitem{francois}F. G\'elis, these proceedings.
\bibitem{sinha98}S. Sarkar, J. Alam, P. Roy, A. K. Dutt-Mazumder, B.
Dutta-Roy, and B. Sinha, Nucl. Phys. {\bf A634} (1998) 206.
\bibitem{martine}M. Bertrand, MSc Thesis (McGill University, 2001),
unpublished.
\bibitem{a1} L. Xiong, E. Shuryak, and G. Brown, Phys. Rev. D {\bf 46}
(1992) 3798; C. Song, Phys. Rev. C {\bf 47} (1993) 2861; C. Song, C. M.
Ko, and C. Gale, Phys. Rev. D {\bf 50} (1994) 1827.
\bibitem{photons}J. Steele, H. Yamagishi, and I. Zahed, Phys. Lett.
{\bf B384} (1996) 255; Miklos A. Halasz, James V.  Steele, Guo-qiang Li, 
Gerald E. Brown, Phys. Rev. C {\bf 58} (1998) 365. 
\bibitem{gg}See, for example, Song  Gao and Charles Gale, Phys. Rev. C
{\bf 57} (1998) 254.
\bibitem{hgr}C. Gale, M. A. Halasz, and R. Rapp, in preparation.
\bibitem{h3}M. Masera, Nucl. Phys. {\bf A590} (1995) 93c.
\bibitem{naimr}P. Bordalo, Nucl. Phys. {\bf A661} (1999) 538c.
\bibitem{pdg}Particle Data Group, Eur. Phys. J. C {\bf 15} (2000) 1.
\bibitem{dol91}S. I. Dolinsky {\it et al.}, Phys. Rep. {\bf 202}
(1991) 99.
\bibitem{ioulia}I. Kvasnikova, PhD Thesis (McGill University, 2001),
unpublished.
\bibitem{ligale}G. Q. Li and C. Gale, Phys. Rev. Lett. {\bf 81} (1998)
1572; Phys. Rev. C {\bf 58} (1998) 2914.
\bibitem{argus}N. Albrecht {\it et al.}, Phys. Lett. {\bf B185} (1987)
223.
\bibitem{rs}Ralf Rapp and Edward Shuryak, Phys. Lett. {\bf B473} 
(2000) 13.  
\bibitem{kgss}D. K. Srivastava {\it et al.}, these proceedings.
\bibitem{kgss2}I. Kvasnikova, C. Gale, D. K. Srivastava, and B. Sinha,
in preparation.
\bibitem{carlos}C. Louren\c{c}o, these proceedings. 
\bibitem{kampfer}B. Kampfer {\it et al.}, these proceedings.
 
\end{thebibliography}
\end{document}